\documentclass[12pt]{article}

\usepackage{epsfig}

\newcommand{\eq}{\begin{equation}}
\newcommand{\eqx}{\end{equation}}
\newcommand{\eqn}{\begin{eqnarray}}
\newcommand{\eqnx}{\end{eqnarray}}

\newcommand{\cor}[1]{\left\langle{#1}\right\rangle}

\begin{document}

\title{Towards the matrix model of M-theory on a lattice}

\author{R.A. Janik$^{a,b}$  and J. Wosiek$^b$ \\ \\
$^a$Service de Physique Theorique  CEA-Saclay \\ F-91191
Gif-sur-Yvette Cedex, France\\
$^b$M.Smoluchowski Institute of Physics, Jagellonian University\\ Reymonta
4, 30-059 Cracow, Poland}

\maketitle

\begin{abstract}
The Wilson discretization of the dimensionally reduced supersymmetric
Yang-Mills theory is constructed. This gives a lattice version
of the matrix model of M-theory. An SU(2) model is studied numerically
in the quenched approximation for D=4. The system shows canonical scaling
in the continuum limit. 
A clear signal for a prototype of the 
``black hole to strings'' phase transition is found. The pseudocritical
temperature is determined and the temperature dependence of the total size
of the system is measured in both phases. Further applications are outlined. 
\end{abstract}
 
\vspace*{1cm}
\noindent TPJU-2/00 \newline
March 2000 \newline
hep-th/0003121
\newpage

1. {\em Introduction and lattice formulation.} The matrix model 
of M-theory provides a qualitative 
description of many essential properties of the final unifying theory
\cite{BFSS}. The interest in studying quantitatively its properties is
particularly enhanced due to a rich spectrum of predictions from
supergravity, for example the thermodynamical properties extracted
from black hole solutions, for the system of D0 branes described by this
SYM quantum mechanics \cite{MAR,POLCH}. 

The complete solution of the above matrix model is  
not known even though it is much simpler than a conventional field theory.
In this letter we construct the Wilson discretization of the model,
and study its yet simpler version with the well developed lattice methods.
In particular, we find the onset of a prototype
black hole to strings
phase transition and determine quantitatively 
some of its properties. Of course the true black hole 
phase of the full D0 brane system is much more complex.
It is therefore quite appealing that the simplified
model considered in this exploratory study
reveals an important part of the phase structure.

Many other problems can be attacked within the 
present approach opening a new area of exciting applications.  
To our knowledge, this is the first study of the M-theory related 
quantum mechanics on a lattice. The zero 
dimensional system has been considered recently in an 
attempt to understand the compactification of higher dimensions 
\cite{JAP,KOP}. Monte Carlo and analytical study of the relevant SU(N) 
integrals have been reported in \cite{KRAU}.

      We begin with the Banks, Fishler, Susskind and Shenker (BFSS)
proposition to use the dimensionally reduced
SUSY YM theory in $D$=10 dimensions  as a model for the 
relevant degrees of freedom of M-theory. For general $D$ the action reads
\cite{UPP}
\eq
   S=\int dt \left({1\over 2} \mbox{\rm Tr} F_{\mu\nu}(t)^2
                     +\bar\Psi^a(t){\cal D}\Psi^a(t) \right). \label{QM}
\eqx
In the process of dimensional reduction all fields are assumed
to be independent of the space variables $x_i$, $i=1 \ldots D-1$ . Consequently
all space derivatives in the field tensor $F_{\mu\nu}$ and in the Dirac
operator ${\cal D}$ vanish ($\partial_i\rightarrow 0$), and (\ref{QM})
describes supersymmetric quantum mechanics
of $D-1$ bosons and their fermionic partners. The temporal components of
the gauge fields are nondynamical and serve to impose Gauss law constraints.
The original $D$ dimensional theory is 
supersymmetric at the classical level only in $D$=2, 4, 6 and 10 dimensions, 
where
appropriate (Majorana, Weyl or both) conditions are imposed \cite{BRINK}.
Fermionic fields $\Psi^a(t)$ belong to the adjoint representation
of the gauge group $SU(N)$, $ a=1\dots N^2-1$. Finally, the BFSS
proposal requires $N\rightarrow\infty$ since this variable corresponds
to the eleventh component of the momentum in the infinite momentum frame
where the original theory is considered.

      We propose to study the above system with methods of the Lattice Field 
Theory. To this end consider $D$ dimensional hypercubic lattice 
$N_1\times\dots\times N_D$ reduced in all space directions to $N_i=1$,
$i=1\dots D-1$. 
Gauge and fermionic variables are assigned to links and sites of the new 
elongated lattice in the standard manner. The gauge part of the action 
reads
\eq
S_G=-\beta
\sum_{m=1}^{N_t} \sum_{\mu>\nu}
{1\over N} Re( \mbox{\rm Tr} \, U_{\mu\nu}(m) ),
\label{SG}
\eqx 
with 
\eq
\beta=2N/a^3 g^2,  \label{beta} 
\eqx
and $U_{\mu\nu}(m)= U_{\nu}^{\dagger}(m)U_{\mu}^{\dagger}(m+\nu)
     U_{\nu}(m+\mu)U_{\mu}(m) $, 
$U_{\mu}(m)=\exp{(iagA_{\mu}(a m))}$, where $a$ denotes the
lattice constant and $g$ is the gauge coupling in one dimension.
The integer time coordinate along the lattice is $m$.
Periodic boundary conditions $U_{\mu}(m+\nu)=U_{\mu}(m)$, $\nu=1\ldots
D-1$,
guarantee that Wilson plaquettes $U_{\mu\nu}$ tend,
in the classical continuum limit, to the 
appropriate components $F_{\mu\nu}$ with space derivatives absent. 
In this formulation the projection on gauge invariant states 
is naturally implemented.

    Discretization of the Dirac operator is analogous to the now standard
construction of the supersymmetric Yang-Mills theories on 
a lattice \cite{MM}. We do not address here important, 
and specific for $D$=10, questions 
of Euclidean formulation for the fermionic degrees of freedom already
discussed in \cite{KRAU} and Weyl projection on the lattice \cite{LU2}. 
Due to the $N_i=1$ periodicity all hopping terms  along the space 
directions collapse into the diagonal blocks of the fermionic matrix.
Then one faces the problem of evaluating fermionic determinant or pfaffian 
for Majorana constraint. For one dimensional system 
the fermionic matrix is effectively block three diagonal. 
This allows for important numerical simplifications. For example, we 
have developed an algorithm which reduces the computational effort
of the exact evaluation of the pfaffian of the antisymmetric fermionic
matrix from $O(V^3)$ to $O(V)$, $V$ being the volume of the system. 
Even with this improvement, however, lattice simulations with dynamical
fermions are much more time consuming than the pure gauge computations. 

    On the other hand experience in
lattice QCD shows that the effect of dynamical fermions is mostly 
accounted for by a shift of the coupling constant while neglecting
the functional dependence of the fermionic determinant \cite{DON}.
Therefore, our first simplifications is to study (\ref{SG}) in the
quenched approximation. Indeed,
as is seen below some essential features ({\em e.g.} the general 
phase structure) of the model are preserved.

       Next simplification concerns the large $N$ limit. Again,
results of lattice simulations for QCD strongly suggest that 't Hooft
choice of the coupling constant, $\lambda=g^2 N$, takes into account 
main large $N$ effects. In fact 
it was found that even results for $SU(2)$ gauge group are not far from 
those with higher $N$ \cite{TEP}. We therefore propose to study systematically
the $N$ dependence of some features of the model (\ref{SG}), beginning with
$N=2$.

     Finally, present simulations are done for $D=4$. Although
the ultimate goal is $D=10$, we expect that the {\em quenched}
approximation, which we study numerically here,
behaves smoothly with $D$ \footnote{This is also confirmed by 
recent analytical results \protect{\cite{KAB}} }. 
Although this will be different in the full unquenched simulation, we
have decided to study first the properties of the simpler system with
$D=4$.  

Needless to say, one should gradually remove the above approximations 
in the forthcoming computations. This is especially important for studying 
the low temperature phase where the supersymmetry restoration may be
essential.  
 
\vspace*{.3cm} 

\noindent 2. {\em Results.} 
One of the most exciting feature of the new theory is the explanation
of the Bekenstein-Hawking entropy puzzle in terms of the microscopic degrees
of freedom of the elementary strings/branes \cite{STRO}. In particular
the theory 
predicts existence of the phase transition at which a black hole
dissolves into its elementary constituents \cite{HOR,MAR}. Confronting
this property with the predictions of the matrix model would provide
an important test of the BFSS conjecture. 
Moreover, lattice study of QCD
at finite temperature show that  the very fact of the
existence of the phase transition is  not sensitive to quenching.
Therefore, as a first application of our construction, we have chosen to study
the phase structure of the system (\ref{SG}). Obviously, the one dimensional
system with local interactions cannot have any phase transition for 
finite N, but just a crossover between two types of the behaviour. 
However for infinite $N$ a sharp 
transition may occur \cite{GW}. 
Thus the number of colours plays 
a role similar to a volume in statistical systems. Consequently,
we expect some signatures of the phase change for finite and even small $N$.
Subsequent simulations for larger N would provide 
more information for the quantitative ({\em e.g.} finite size scaling)
analysis.

As an order parameter we choose the distribution of the Polyakov line
\eq
    P={1\over N}  \mbox{\rm Tr} \left( \prod_{m=1}^{N_t} U_D(m)
\right).  \label{poly} 
\eqx
Similarly to lattice QCD, symmetric concentration of the eigenvalues
around 0 indicates a low temperature phase (which would have the
interpretation of a black hole phase in the full model) where 
$\cor{P}\sim 0$,  while clustering around $\pm 1$ (for SU(2)) is
characteristic of the high temperature (elementary excitations) phase. 
 
A sample of results for different $N_t (\equiv N_D)$ and $\beta$ is
shown in Fig. 1. 
Indeed, for each $N_t$, we see a definite change of the shape with $\beta$.
This is the first result: the system (\ref{SG}) shows unambigously
the onset of the phase change, even in the quenched approximation
and for $N=2$. 

  \begin{table}    
  \begin{center}
   \begin{tabular}{ccc} \hline\hline
   $N_t$ &  $\beta_{low}$ & $\beta_{up}$  \\
   \hline
  2 &  1.25 & 1.5 \\
  3 &  3.5  & 5.0 \\
  4 &  8.0  & 16.0  \\ 
  5 & 15.0  & 40.0  \\ 
  \hline
\multicolumn{3}{c}{$fit:\;\;\;\;$     $\beta_c=\alpha N_t^{\gamma}$ }\\
\hline
 $\chi^2/NDF$    &  $ \alpha $  & $\gamma$ \\
  $.55/2 $       & $.17\pm .05$ & $3.02\pm .33$ \\
   \hline\hline  
   \end{tabular}
  \end{center}
\caption{Estimated location of the transition region
$\beta_c\in(\beta_{low},\beta_{up})$ 
for different lattice sizes $N_t$ and results of the power fit.}
\end{table}

Second, the dependence of the  
pseudocritical temperature $\beta_c$ on the time extent $N_t$ is
consistent with the  
continuum limit expectations $T_c \sim (g^2 N)^{1/3}$ \cite{KAB}.
Indeed, the temperature of a system is given by $T=1/(a N_t)$ .
Together with (\ref{beta}) these relations imply $\beta_c \sim N_t^3$.
 The estimates for $\beta$ intervals where the 
change of phases occur are presented in Table 1 for several lattice 
sizes $N_t$. Results of the 
power law fit are also quoted. A good quality of the fit and the agreement with
the canonical exponent, $\gamma=3$, is encouraging. Simultaneously, we
obtain the proportionality coefficient $\alpha$ which translates into
\eq
  T_c=({\alpha\over 2N^2})^{1/3}(g^2 N)^{1/3}=(.28\pm.03) (g^2 N)^{1/3}. \label{TC}
\eqx
To summarize this point: the observed dependence of $\beta_c$ on $N_t$ 
agrees with the canonical scaling
expectations for the one dimensional system, and indicates
the finite value of the transition temperature in the continuum.
Moreover, the coefficient in the continuum relation (\ref{TC})
has been determined for the first time. Only proportionality of the two
scales has been considered until now \cite{HOR,MAR,KAB}. Since both the 
pseudocritical temperature and $\alpha$ depend in general on $N$, it is 
important to repeat similar analysis for higher gauge groups.

    Next we study the temperature dependence of the total size of the system 
$R^2=g^2\sum_a (A_i^a)^2$ \cite{KAB}. We define for $SU(2)$ 
\eq
    \cor{R^2}\equiv \left(4-\cor{(\mbox{\rm Tr}(U_s))^2}\right)/a^2,
\label{rms} 
\eqx
 where $U_s$ is any space link. 
Due to the periodicity ($N_s=1$) in space (\ref{rms}) is  
gauge invariant.
The space links $U_s$ are the remnants of the torelon observables
well known in lattice QCD \cite{MICH}.

One dimensional Yang-Mills coupling $g$ provides a single scale for all
continuum observables similarly to $\Lambda_{QCD}$ in four dimensions.
In the following all dimensional quantities quoted in units of $g^{2/3}$
are denoted by a tilde.

Even though the 
quantum mechanical system (\ref{SG}) is much simpler than
the full $D$-dimensional field theory, extracting the continuum limit
of the lattice formulation (\ref{SG}) may be a nontrivial task. For
example, the above limit contains the complete information about both 
the perturbative weak coupling and nonperturbative strong
 coupling regimes in the continuum. Technically, relation
 (\ref{beta}) implies that a reasonably small lattice constant, $a$ 
requires simulation with a very large coupling $\beta$.
In addition,  the one 
dimensional systems are harder to thermalize. All this poses an 
interesting challenge in constructing new algorithms suitable
for this problem. Some of such alghorithms are under development
and will be discussed elsewhere. Here we use mostly the standard
local Metropolis update. To overcome the critical slowing down we
simply increase the number of 
thermalization and decorrelation sweeps with $\beta$, until results become
 independent of the starting configuration. This turned out to be 
in accord with the dynamical exponent $z=2$. 
For example when running at $\tilde{a}=1.0$  we used
5000 thermalization and 50 decorrelation sweeps, while for $\tilde{a}=0.1$
about $10^6$ thermalization and 5000 decorrelation sweeps were required.
One of the new algorithms mentioned above is the SU(2) heat bath
designed for an update 
of the space-space plaquettes in (\ref{SG}) which contain twice the same link.
Current version is effective only for $\beta < 64 $. Results obtained
with the new heat bath and the standard Metropolis agree within
statistical errors. To check independently the performance of the
Metropolis algorithm for higher $\beta$ we have also monitored the
correlation length in the torelon channel at zero ({\em i.e.} low)
temperature. It reveals the expected canonical scaling with $a$. 

Fig. 2 shows the 
dependence of $\tilde{R}^2$ on $\tilde{a}$, for several values of the 
temperature $\tilde{T}$. 
MC results depend smoothly on $a$, at fixed T, which confirms
the existence of the continuum limit (\ref{rms}). The $a$ dependence
is clearly  
different in low and high temperature regions. For $.1<\tilde{T}<.3$,
$\cor{\tilde{R}^2}$ is 
practically independent of $\tilde{a}$
and points for different (but small) $\tilde{T}$ collapse on the same line. 
For higher $\tilde{T}$ quadratic minimum
at $\tilde{a}=0$ develops and shrinks with the further increase of the temperature.
For $\tilde{T}>1.5 $ simulations for smaller $\tilde{a}$ are
required in order to see this structure and determine the continuum limit. We
have also extracted $\cor{R^2}$ from another lattice observable
$|Tr(U_s)|$ with practically the same results.

Fig. 3 shows the size of a system extrapolated
to $a=0$ as a function of the temperature. 
Both quadratic and quartic fits of $a$ dependence were used to perform the
 extrapolation \cite{SYM}.
 We have also checked the stability
of quadratic fits with respect to removing one or two data points with
smallest $a$ (highest $\beta$). Results of the extrapolation were
stable with respect to all these variations. 
Small systematic shifts are included in the errors displayed in Fig. 3.
The location of the transition region
is in rough agreement
with the estimate (\ref{TC}) of the pseudocritical temperature
$\tilde{T_c}=0.35\pm0.04$ 
\footnote{The pseudocritical temperatures determined from
different observables can be different.}. Again, it is evident that the 
system is indeed different in the two regimes. Moreover, our results
agree qualitatively 
with the analytical prediction obtained by solving a gap equation 
in the infinite N limit \cite{KAB}. The latter gives a temperature 
independent constant at low temperatures and the classical $T^{1/2}$ 
growth for high temperatures. We have also found a reasonable agreement
with a simple mean field model for $SU(2)$ with the gauge
projection\footnote{To be discussed in detail elsewhere.}. 
 As expected the model does not have a phase transition,
but shows a smooth crossover located as in Fig.3. The constant value
for $\cor{R^2}$ is 
satisfactorily reproduced in the low
temperature, vacuum driven region. At higher temperatures the model predicts
intermediate linear, albeit weaker than MC, behaviour which
asymptotically turns over into $T^{1/2}$ as in the infinite $N$ case.

\noindent {\em 3. Conclusions.}
We have constructed the matrix model of M-theory on a lattice in
$D$=2,4,6 and 10 dimensions. The resulting system corresponds to the
supersymmetric formulation of Yang-Mills theory on the asymmetric
$D$-dimensional lattice with all $D-1$ space extensions $N_s=1$.
The new construction was tested in the quenched approximation for $D$=4. 
In particular, we have found the onset of 
a black hole to strings transition even for the $SU(2)$ 
gauge group. 
The pseudocritical temperature was determined. The size of the
system was also measured at different temperatures and lattice cut-offs.
It shows the expected canonical scaling. After extrapolation to the continuum
limit it confirms the existence of the two phases
and agrees qualitatively with the mean field calculations.

      A host of new applications can follow. On the technical side,
new algorithms are required to reduce the critical slowing down at 
very large values of the lattice coupling. Such studies have 
already begun. Including dynamical fermions is facilitated by
the linear nature of the system and may lead to more efficient
fermionic algorithms. Certainly the issue
of dynamical fermions is very important especially in the low 
temperature phase since one
expects that supersymmetry should be broken only in a minimal fashion there. 
With dynamical fermions in $D$=10 one may have 
to use the recently proposed chiral formulation \cite{LU2}. 
On the other hand for the reduced system the task may be 
simpler than e.g. for QCD. It would also be very interesting to apply
analytical methods developed in \cite{LU1,VABA}.
Incidentally, a merit of the present approach is the possibility to
draw from the expertise, techniques and algorithms developed in the
lattice community.

A systematic study of the model for higher $N$ would allow finite size
analysis and determine more detailed characteristics of the
transition. In particular it would be interesting to check if the
``soft'' dependence on $N$ observed for 3D and 4D $SU(N)$ lattice YM
\cite{TEP}, persists in the SYM quantum mechanical model.

Finally, one of the ultimate physical goals would be to study the
thermodynamics of the black-hole phase in the full D=10 model and
verify existence of the rich phase structure predicted by the string/M
theory \cite{MAR}. This would also provide a possible nontrivial
quantitative test of (a version of) the AdS/CFT correspondence at
strong coupling not protected by any nonrenormalization theorems
\cite{POLCH,YON}. 

Last but not least, many other problems inspired by the BFSS
conjecture can be quantitatively studied.

\vspace*{0.5cm}
 
\noindent {\em Acknowledgements.} This work is  supported by the
Polish Committee for Scientific Research under grants no. PB
2P03B00814 and PB 2P03B01917.

\section*{Figures}

\begin{figure}[htb]
%\vspace{-10cm}\hspace*{-5cm}
%\framebox[100mm]{
%\rule[-21mm]{0mm}{43mm}
\epsfig{width=12cm,file=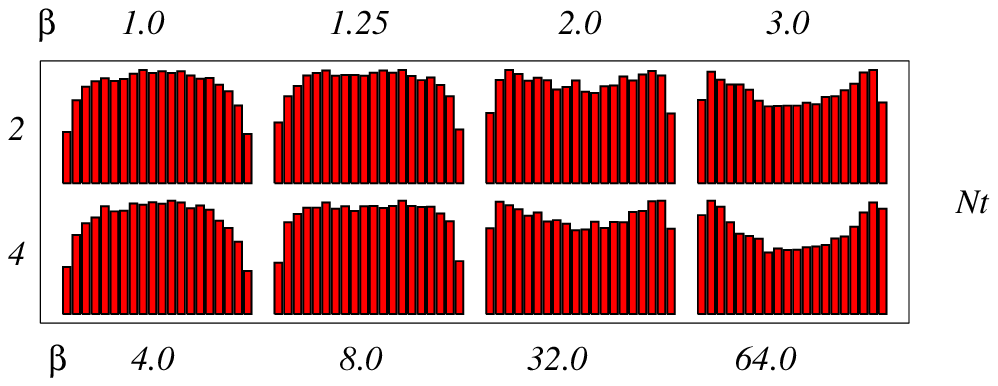}
% }
% \vspace*{-10cm}
\caption{Distribution of the Polyakov line (\ref{poly}), $-1<P<1$, for different $\beta$ and $N_t$.
Note different $\beta$ range for different $N_t$.  
}
\label{fig:f1}
\end{figure}

\begin{figure}[htb]
%\vspace{-10cm}\hspace*{-5cm}
%\framebox[55mm]{\rule[-21mm]{0mm}{43mm}}
\epsfig{width=12cm,file=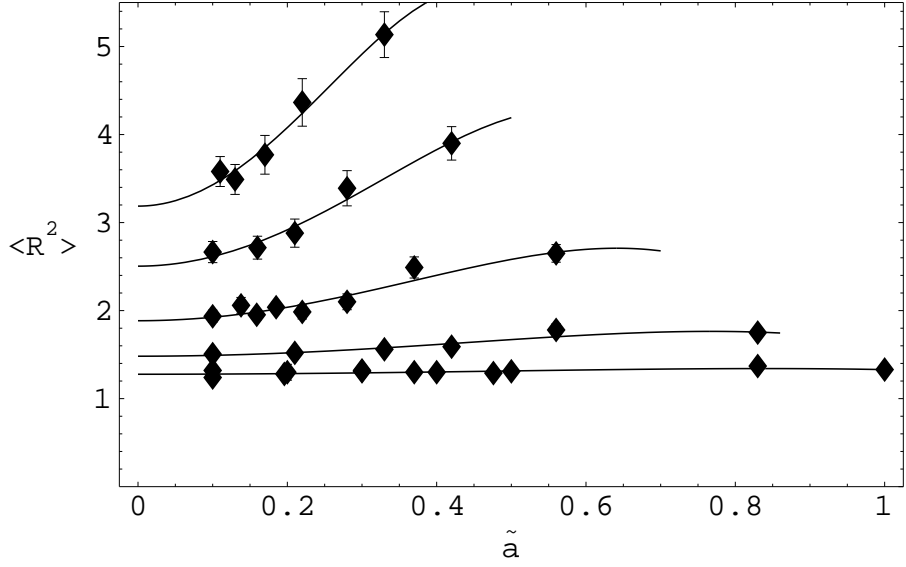}
%\vspace*{-10cm}
\caption{ Dependence of the total size of the system on $a$, for
 $\tilde{T}=.1-.3,.6,.9,1.2$ and $1.5$ (upwards) 
in units of $g^{2/3}$. Quartic fits are represented by the solid lines. }
\label{fig:f2}
\end{figure}

\begin{figure}[htb]
%\vspace{-10cm}\hspace*{-5cm}
%\framebox[55mm]{\rule[-21mm]{0mm}{43mm}}
\epsfig{width=12cm,file=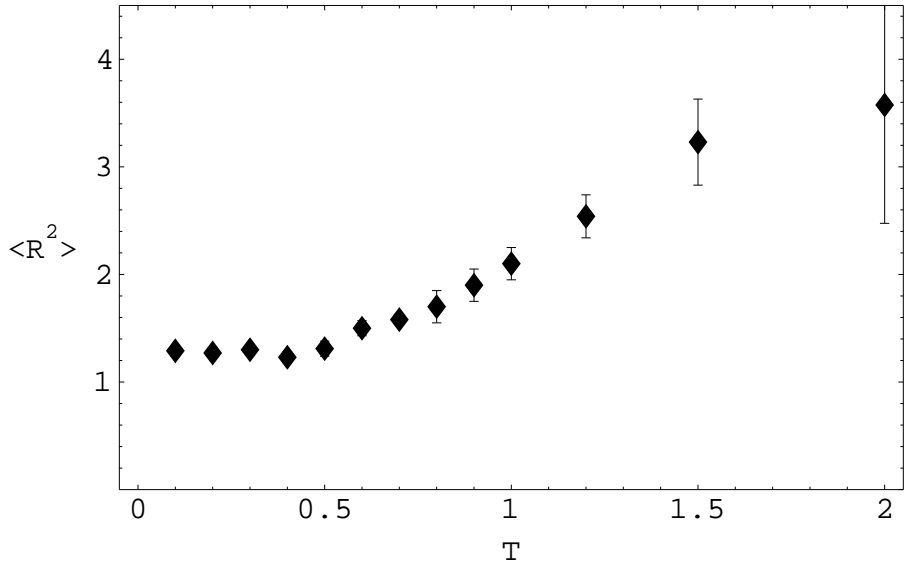}
%\vspace*{-10cm}
\caption{ Size of the system (\ref{rms}) extrapolated to the continuum,
 as a function of the temperature.}
\label{fig:f3}
\end{figure}

\end{document}